\documentclass[preprint,aps,showpacs]{revtex4}
\usepackage{graphicx}
\begin{document}
 
\title{Coulomb excitation at intermediate energies}
\author{H. Esbensen}
\affiliation{Physics Division, Argonne National Laboratory, Argonne, Illinois 60439, USA} 
\date{\today}
 
\begin{abstract}
Straight line trajectories are commonly used in semi-classical 
calculations of the first-order Coulomb excitation cross section
at intermediate energies, and simple corrections are often made
for the distortion of the trajectories that is caused by the Coulomb field.
These approximations are tested by comparing to numerical calculations 
that use exact Coulomb trajectories.
In this paper a model is devised for including relativistic effects 
in the calculations.
It converges at high energies towards the relativistic 
straight-line trajectory approximation and approaches 
the non-relativistic Coulomb trajectory calculation at low energies.
The model is tested against a number of measurements and analyses that 
have been performed at beam energies between 30 and 70 MeV/nucleon,
primarily of quadrupole excitations.
Remarkably good agreement is achieved with the previous analyses, 
and good agreement is also achieved in the few cases, 
where the B(E$\lambda$) value is known from other methods. 
The magnitudes of the relativistic and Coulomb distortion effects 
are discussed.
\end{abstract}
 
\maketitle

\section{Introduction}

The first-order excitation of a nucleus, induced by the Coulomb field 
from another nucleus, has been discussed in detail in Ref.  \cite{AW-book}.
The semi-classical perturbation theory developed there was based on a 
classical, non-relativistic Coulomb orbit for the relative motion of the 
projectile and target nuclei. A relativistic description was later  
developed but it was restricted to straight-line trajectories \cite{AW-NPA}. 
This leaves a gap in the theoretical description at intermediate energies
(say, at 20-200 MeV/nucleon), 
where both relativistic effects and Coulomb distortions of the trajectory 
can be important.  This is very unfortunate because many experiments have
been performed in recent years at intermediate-energy, radioactive beam 
facilities. 
These include Coulomb dissociation experiments, for example, of $^8$B, 
$^{11}$Be and $^{11}$Li, and Coulomb excitation measurements in
inverse kinematics, mostly of the lowest $2^+$ state in light and 
medium heavy nuclei \cite{glass}.

The purpose of this paper is to devise a model that interpolates 
smoothly between the non-relativistic description of the Coulomb excitation
which is based on Coulomb trajectories \cite{AW-book} and the relativistic 
description which is based on straight-line trajectories \cite{AW-NPA}.
To develop an exact theory that contains the two extreme descriptions 
as limits is in general a difficult task. It has been studied by
Bertulani et al.  \cite{aleixo,bert68}, who considered the effects of 
retardation explicitly for a Coulomb trajectory.
The method proposed here is much simpler but it is sufficiently accurate for 
analyzing the data that have been obtained at intermediate energies. 
The point is that the experimental uncertainties are often large, typically 
of the order of 10\%, and there are also theoretical uncertainties, which 
can distort the analysis of data, for example, the influence of nuclear and 
higher-order processes.

The accuracy of the proposed model of the intermediate-energy Coulomb 
excitation is tested by analyzing the results obtained with the commonly 
used `Coulomb corrected' straight-line trajectory method, which originally 
was proposed by Winther and Alder \cite{AW-NPA}.
Another test is to use B(E2) values that are known 
from 
other measurements (for example, of the life-time) and compare the calculated 
cross sections to measurements that have been performed at intermediate
energies.

There has recently been a debate in the literature \cite{bertpl,scheit} 
about the validity of the analyses of Coulomb excitation experiments that 
have been performed in the past. It turned out to be caused by a 
misunderstanding of the experimental conditions, as pointed out in 
Ref. \cite{scheit}.
However, independent of the controversy, it was claimed that corrections 
to the low-energy and high-energy theories of Coulomb excitation could be 
as large as 20\% or 30\% at intermediate energies \cite{bert68,bertpl}.
In this paper 
it will be shown how this large uncertainty can be brought under
control and reduced to only a few percent.
  
The semi-classical theory of the non-relativistic Coulomb excitation is 
summarized in sections II to III. The expressions that are obtained from 
straight-line and Coulomb trajectories, respectively, are compared 
in section IV.
The analytic expression for the relativistic Coulomb excitation in the
straight line trajectory approximation is quoted in Section V, and it 
is used to devise a model which includes the combined effect of relativity 
and Coulomb trajectories.
The model is tested in section VI against measurements and other
calculations, and section VII contains the conclusions.

\section{Non-relativistic Coulomb excitation}

The semi-classical, non-relativistic description of Coulomb excitation 
\cite{AW-book} is summarized in the following.
Thus we consider a target nucleus with atomic number $Z_2$ which is 
being excited by the Coulomb field from a projectile nucleus
with atomic number $Z_1$.
It is assumed that the projectile and target do not overlap during the 
collision. We can therefore use the so-called far-field approximation 
which assumes that the intrinsic coordinate $r$ of the target nucleus 
is smaller that the distance $R(t)$ between 
projectile and target.  The first-order amplitude for the electric 
excitation of the target nucleus, from an initial state $|i>$ to  
a final state $|f>$, is given by the multipole expansion \cite{AW-book},
\begin{equation}
a_{fi} = \frac{Z_1e^2}{i\hbar} 
\sum_{\lambda\mu} \sqrt{\frac{4\pi}{2\lambda+1}}
S_{\lambda\mu}(\omega) 
<f|M_{\lambda\mu}^*|i>,
\end{equation}
where $\hbar\omega=\Delta E_{fi}$ is the excitation energy, 
$M_{\lambda\mu}$ is the multipole operator 
$r^\lambda Y_{\lambda\mu}({\hat r})$, and 
\begin{equation}
S_{\lambda\mu}(\omega) = 
\sqrt{\frac{4\pi}{2\lambda+1}}
\int_{-\infty}^\infty dt \ e^{i\omega t} \
\frac{Y_{\lambda\mu}({\hat R}(t))}{R(t)^{\lambda+1}}
= \int_{-\infty}^{\infty} dt \
e^{i\omega t} \ \frac{d_{\mu 0}^\lambda({\beta}) e^{i\mu\phi}}
{(R(t))^{\lambda+1}}
\label{orbint}
\end{equation}
is the so-called orbital integral. It depends on the orbit 
${\vec R}(t)$ of the projectile with respect to the target nucleus.
The unit vector ${\hat R} = {\vec R}/R$ is expressed in terms of 
the spherical coordinates $(\beta,\phi)$ in the last expression. 
It is noted that this definition of the orbital integral, 
Eq.  (\ref{orbint}), differs by the factor $\sqrt{4\pi /(2\lambda+1)}$
from the convention used in Ref. \cite{AW-book}. 

The calculation of the orbital integrals is discussed in the
following sections for two choices of the coordinate system.
The first choice, system A, is the most convenient for numerical
calculations that are based on a 
Coulomb orbit \cite{AW-book}. 
The second choice, referred to as system H, is more convenient 
at high energies, where a straight line trajectory becomes an accurate 
approximation and relativistic effects can be treated exactly \cite{AW-NPA}.
The transformation between the two representations will be discussed 
in order to be able to compare the two extreme methods and devise a 
model that interpolates smoothly between them.

\subsection{Cross sections}

The Coulomb excitation cross section is calculated as the product of the 
first-order excitation probability $P_{fi}$ and the elastic Rutherford 
cross section $(d\sigma/d\Omega)_R$,
\begin{equation}
\Bigl(\frac{d\sigma}{d\Omega}\Bigr)_{fi} = 
P_{fi} \ \Bigl(\frac{d\sigma}{d\Omega}\Bigr)_{\rm R}.
\label{crosss}
\end{equation}
The Rutherford cross section can be obtained from the Rutherford
scattering formula, $\tan(\theta/2)=a/b$, where $b$ is the impact 
parameter and 
\begin{equation}
a = \frac{Z_1Z_2e^2}{M_0v^2}, 
\label{headon}
\end{equation}
is half the of distance of closest approach in a head-on collision,
and $M_0$ is the reduced mass, $M_0=M_1M_2/(M_1+M_2)$.
The Rutherford cross section is
\begin{equation}
\Bigl(\frac{d\sigma}{d\Omega}\Bigr)_{\rm R} =  
\frac{2\pi b db}{d\Omega}
=\pi a^2 \ \frac{d}{d\Omega} 
\Bigl(\frac{1}{\sin^2(\theta/2)}\Bigr).
\end{equation}
 
The average excitation probability obtained from Eq. (1) 
is the average over the initial magnetic sub-states $M_i$  
and the sum over the final m-sub-states $M_f$,
$$
P_{fi} = \frac{4\pi Z_1^2 e^4}{(2\lambda+1)\hbar^2} \ 
\sum_{\mu} |S_{\lambda\mu}|^2 \
\frac{1}{2I_i+1} 
\sum_{M_iM_f} |\langle I_fM_f |M_{\lambda\mu}^*| I_iM_i\rangle|^2.
$$
The last sum divided by $(2I_i+1)$ is equal to $(2\lambda+1)^{-1}$ 
times the multipole strength (or reduced transition probability)
for the excitation, i.~e.,
\begin{equation}
P_{fi} = 4\pi 
\ \Bigl(\frac{Z_1 e^2}{(2\lambda+1)\hbar}\Bigr)^2 
\ B(E\lambda) \ \sum_{\mu} |S_{\lambda\mu}|^2. 
\label{excprob}
\end{equation}

\section{Orbital integrals in coordinate system A}

In the coordinate system denoted by A (c.f. Ref. \cite{AW-book}), 
the z-axis is perpendicular to the scattering plane so the orbit 
has the form ${\vec R}(t)=(x(t),y(t),0)$. The angle $\beta$ is 
fixed at $\pi/2$, and the orbital integral (2) can be written as 

\begin{equation}
S_{\lambda\mu}^A = d_{\mu 0}^\lambda({\pi\over 2})
\int_{-\infty}^\infty dt \ e^{i\omega t} \
{(x(t)+iy(t))^\mu\over (R(t))^{\lambda+\mu+1}}.
\label{slma}
\end{equation}
The factor $d_{\mu 0}^\lambda(\pi/2)$ ensures that the orbital 
integral is non-zero only when $\lambda+\mu$ is even.

In the coordinate system A one chooses the x-axis as the symmetry 
axis of the Coulomb orbit so that $x(-t)=x(t)$ and $y(-t)=-y(t)$. 
From this it follows that 
$S_{\lambda\mu}^A = \Bigl(S_{\lambda\mu}^A\Bigr)^*$ is a real 
quantity.  To calculate $S_{\lambda-\mu}^A$ when $\mu>0$ 
one can use the expression
\begin{equation}
S_{\lambda-\mu}^A = d_{\mu 0}^\lambda({\pi\over 2})
\int_{-\infty}^\infty dt \ e^{i\omega t} \
{(x(t)-iy(t))^\mu\over (R(t))^{\lambda+\mu+1}}.
\end{equation}
Below we discuss the calculation for a Coulomb trajectory and 
compare it to the result of the straight line approximation.

\subsection{Coulomb trajectories}

To evaluate the orbital integrals numerically for a Coulomb orbit
one makes use of a dimensionless time-variable $w$ 
(see Ref. \cite{AW-book}) so that
\begin{equation}
R(t) = a[\epsilon \cosh(w) +1], \ \ \
t={a\over v}[\epsilon \sinh(w) + w],
\label{coultr}
\end{equation}
where $a$ is defined in Eq. (\ref{headon}) and
$\epsilon$ is the eccentricity of the orbit,
which can be expressed in terms of the impact parameter $b$ or
the scattering angle $\theta$ in the center of mass system as follows,
\begin{equation}
\epsilon = \sqrt{1+(b/a)^2} = \frac{1}{\sin(\theta/2)}.
\label{eps}
\end{equation}

Inserting the Cartesian coordinates of the orbit (see \cite{AW-book}):
\begin{equation}
x = a[\cosh(w)+\epsilon], \ \ 
y = a\sqrt{\epsilon^2-1} \sinh(w), \ \ z=0,
\end{equation}
into Eq. (\ref{slma}) one obtains,
\begin{equation}
\label{cltr}
S_{\lambda\mu}^A = {1\over v a^\lambda}
d_{\mu 0}^\lambda\bigl({\pi\over 2}\bigr) \ I_{\lambda\mu},
\end{equation}
where
\begin{equation}
\label{ILM}
I_{\lambda\mu} = \int_{-\infty}^\infty dw 
\ \exp[i\xi_a(\epsilon \sinh(w) +w)]
{[\cosh(w)+\epsilon+i\sqrt{\epsilon^2-1}\sinh(w)]^\mu\over
(\epsilon \cosh(w)+1)^{\lambda+\mu}},
\end{equation}
and $\xi_a=\omega a/v$. 
Properties and even tabulations of $I_{\lambda\mu}$ are given in 
Ref. \cite{AW56}.  It will be calculated using a simple numerical 
integration with respect to $w$ over the finite interval [-5:5],
and using just a few thousand steps. The accuracy of the numerical 
integration can be tested in the case of a straight-line trajectory 
against the analytic expressions, which are discussed in appendix A.


  
\section{Coordinate system H}

At high energies it is more convenient to use the coordinates system H
where the z-axis is along the beam direction and the x-axis is along 
the impact parameter so that the y-axis is perpendicular 
to the scattering plane.
The coordinates of the trajectory are therefore ${\vec R}(t)$ = 
$(x(t),0,z(t))$, which implies that $Y_{\lambda\mu}({\hat R})$ is real. 
Since $Y_{\lambda\mu}^*=(-1)^\mu Y_{\lambda-\mu}$, it follows directly 
from the definition (2) that the orbital integral in coordinate system H
must have the property,
\begin{equation}
S_{\lambda-\mu}^H(\omega) = (-1)^\mu S_{\lambda\mu}^H(\omega).
\end{equation}

The coordinate system H is convenient at high energies because
the analytic expressions for the orbital integrals that exist for 
straight-line trajectories are relatively simple in this representation 
even when the relativistic effects are included \cite{AW-NPA}. 
This feature will be exploited in section V to devise an expression 
that contains the effects of relativity and the Coulomb distortion of 
the trajectory.  To do that, we will need to transform the orbital 
integrals in coordinate system A to the new coordinate system H.

To go from the H to the A coordinate system consists of a rotation 
of $\pi/2$ around the z-axis, followed by a rotation of $-\pi/2$ around 
the new y-axis, and finally a rotation of $-\pi/2$ around the final z-axis.
The required transformation is therefore
\begin{equation}
S_{\lambda\mu}^H = \sum_{\mu'} i^{\mu-\mu'}
\ d_{\mu\mu'}^\lambda\bigl(-{\pi\over 2}\bigr) 
S_{\lambda\mu'}^A.
\end{equation}
Inserting Eq. (\ref{cltr}) into this transformation we can
write that 
\begin{equation}
S_{\lambda\mu}^H = 
\frac{i^{\lambda+\mu}}{va^\lambda} \ I_{\lambda\mu}^H, \ \ \
{\rm where} \ \ I_{\lambda\mu}^H =
\sum_{\mu'} i^{-(\lambda+\mu')}
\ d_{\mu'\mu}^\lambda\bigl({\pi\over 2}\bigr) 
\ d_{\mu' 0}^\lambda\bigl({\pi\over 2}\bigr)
\ I_{\lambda\mu'}.
\label{ILMH}
\end{equation}

Values of $d_{\mu'\mu}^\lambda(\pi/2)$ can be obtained from appendix 
D of Ref. \cite{AW-book}.  
The explicit expressions one obtains for dipole and quadrupole 
excitations are
\begin{equation}
I_{10}^H = \frac{I_{1-1} - I_{11}}{2}, \ \ \ 
I_{11}^H = \frac{I_{1-1} + I_{11}}{2\sqrt{2}},
\label{ILMH1}
\end{equation}
\begin{equation}
I_{20}^H = \frac{3}{8} \Bigl[ I_{22}+I_{2-2} - \frac{2}{3} I_{20}\Bigr], \ \
I_{21}^H = \frac{1}{2} \sqrt{\frac{3}{8}} 
\Bigl[ I_{2-2}-I_{22}\Bigr], \ \
I_{22}^H = \frac{1}{4} \sqrt{\frac{3}{8}} 
\Bigl[ I_{22}+ I_{2-2} +2I_{20} \Bigr].
\label{ILMH2}
\end{equation}

\subsection{Straight line trajectories}

In the straight-line trajectory approximation, the projectile moves
with constant velocity $v$ along the z-axis at an impact parameter
$b$ with respect to the target, ${\vec R}(t)=(b,0,vt)$. 
The non-relativistic orbital integrals have the analytic form \cite{AW-NPA},
\begin{equation}
{\tilde S}_{\lambda\mu}^H = {2\over v} \ 
{i^{\lambda+\mu}\over \sqrt{(\lambda+\mu)!(\lambda-\mu)!}}
\ \Bigl({\omega\over v}\Bigr)^\lambda \ K_\mu(\xi_b),
\label{strlh} 
\end{equation}
where $\xi_b=\omega b/v$ is the adiabaticity parameter
associated with the impact parameter $b$.
The tilde on ${\tilde S}_{\lambda\mu}^H$ is a reminder of 
the straight-line trajectory approximation, and the superscript 
H refers to the coordinate system used here.
It should be emphasized that the expression Eq. (\ref{strlh})
can be derived by inserting the straight-line approximation,
Eq. (A3), into the transformation, Eq. (\ref{ILMH}).

\subsection{Coulomb trajectories} 

In order to test the accuracy of Eq. (\ref{strlh})
it is useful to write the general Coulomb trajectory result, 
Eq. (\ref{ILMH}), in a similar form, 
\begin{equation}
S_{\lambda\mu}^H = {2\over v}
{i^{\lambda+\mu}\over \sqrt{(\lambda+\mu)!(\lambda-\mu)!}}
\ \Bigl({\omega\over v}\Bigr)^\lambda 
K_{\lambda\mu}^{eff}(b/a,\xi_a),
\label{nonrel}
\end{equation}
where
\begin{equation}
K_{\lambda\mu}^{eff}(b/a,\xi_a) =
{1\over 2} \Bigl({1\over \xi_a}\Bigr)^\lambda \
\sqrt{(\lambda+\mu)!(\lambda-\mu)!} \ 
I_{\lambda\mu}^H,
\label{nrelk}
\end{equation}
and $I_{\lambda\mu}^H$ are the orbital integrals defined in
coordinate system H according to Eq. (\ref{ILMH}). They are
given explicitly by Eqs. (\ref{ILMH1}) and (\ref{ILMH2}) for 
dipole and quadrupole excitations.


One can check that $K_{\lambda\mu}^{eff}$ gives 
the correct modified
Bessel function $K_\mu(\xi)$ when one inserts the straight-line trajectory
results in coordinate system A, Eqs. (A4-A6) of  
appendix A, into the definition (\ref{ILMH}) of $I_{\lambda\mu}^H$.
In the limit: $\xi_a<<1$ and $b/a>>1$, i.~e., at high energies and large
impact parameters, one should recover the result for a straight line,
i.~e. $K_{\lambda\mu}^{eff}(b/a,\xi_a)$ $\rightarrow$ $K_\mu(\xi_b)$.
This convergence will be illustrated in the next section.

To evaluate the excitation probability (\ref{excprob}), 
we need the expression
$$ \sum_{\mu} |S_{\lambda\mu}^H|^2 = 
\frac{4}{v^2 \ b^{2\lambda}} \ F_\lambda(b/a,\xi_a),$$
where
\begin{equation}
F_\lambda(b/a,\xi_a) = \sum_\mu 
\frac{\xi_b^{2\lambda}}{(\lambda+\mu)!(\lambda-\mu)!} \
\Bigr(K_{\lambda\mu}^{eff}(b/a,\xi_a)\Bigr)^2.
\label{fcoul}
\end{equation}
In the straight-line trajectory approximation, where
$K_{\lambda\mu}^{eff}(b/a,\xi_a)$ $\rightarrow$ $K_\mu(\xi_b)$,
one obtains 
\begin{equation}
F_1(\xi_b) = \xi_b^2 \Bigl(K_0^2(\xi_b) + K_1^2(\xi_b)\Bigr),
\ \ \ {\rm and}
\ \ \
F_2(\xi_b) = \frac{\xi_b^4}{12} \
\Bigl(3K_0^2(\xi_b) + 4 K_1^2(\xi_b) + K_2^2(\xi_b)\Bigr).
\label{fstrl}
\end{equation}
for dipole and quadrupole excitations, respectively.
For $\xi_b\rightarrow$ 0 these functions approach the constant values 
$F_1\rightarrow$1 and $F_2\rightarrow$1/3.

\subsection{Comparison of results}

Here the convergence of the Coulomb trajectories calculations
towards the straight-line trajectory calculation is 
illustrated by comparing the functions $F_\lambda(b/a,\xi_a)$ 
defined in Eq. (\ref{fcoul}) to the analytic expressions (\ref{fstrl}) 
for dipole and quadrupole excitations.
The solid curves in Fig. 1 show the results of the straight-line 
trajectory approximation, Eq. (\ref{fstrl}), for dipole ($\lambda=1$) 
and quadrupole ($\lambda=2$) excitations, respectively, as functions 
of the adiabaticity parameter $\xi_b=\omega b/v$. 
The results of the Coulomb trajectory calculations defined in 
Eq. (\ref{fcoul}) are shown by dashed curves for three values of $b/a$. 
It is seen that the Coulomb trajectory results approach the straight-line 
result in a smooth manner for increasing values of $b/a$.

The solid circles in Fig. 1 are the results one obtains by multiplying
the straight-line trajectory expressions, Eq. (\ref{fstrl}), with the 
factor $\exp(-\pi\xi_a)=\exp(-\pi\xi_b a/b)$ for $b/a$=5. 
This correction factor was suggested by Winther and Alder \cite{AW-NPA} 
and it is seen to reproduce the dipole results ($\lambda$=1)
for the Coulomb trajectory with $b/a$=5 fairly well. 
However, it does not work so well for quadrupole excitations ($\lambda$=2).
The problem is that the Coulomb trajectory results depend on $b/a$
for $\xi_b\rightarrow 0$, whereas the simple correction factor
$\exp(-\pi\xi_a)=\exp(-\pi\xi_b a/b)$ is 1 in this limit.

The non-relativistic cross sections one obtains for the dipole excitation 
of the $^{11}$Be $1/2^+$ ground state to the $1/2^-$ excited state 
and for the quadrupole excitation of $^{42}$S to the $2^+$ state
are illustrated in Fig. 2. The cross sections were calculated for a gold 
target with a minimum impact parameter of 14 fm and they are shown as 
functions of the beam energy.  Although these cross sections are referred to
as non-relativistic, it should be emphasized that the velocity that has been
used here was actually obtained from the relativistic expression, Eq. (B1), 
in terms of the beam energy.

The excitation energies and B(E$\lambda$) values that have been used for $^{11}$Be 
and $^{42}$S are shown in Table I, which will be discussed later.
The top dashed curves in Fig. 2 show the straight-line trajectory calculations,
and the thin dashed curves are the same results multiplied by the
factor $\exp(-\pi\xi_a)$.  The solid curves are based on Coulomb trajectories. 
They approach the straight-line trajectory approximation fairly quickly 
for the dipole excitation but somewhat slower for the quadrupole excitation.
It is also seen that the approximation of multiplying the straight
line calculation with the factor $\exp(-\pi\xi_a)$ works quite well
for the dipole excitation when compared with the Coulomb trajectory 
calculation, whereas this approximation is poorer for quadrupole 
excitations.  Other approximations have been applied to correct the 
straight line trajectory approximation for the distortion that is 
caused by a Coulomb trajectory and some of them will be discussed in
section VI.D. 


\section{Relativistic expression} 

The relativistic expressions for the orbital integrals in the straight-line 
trajectory approximation are \cite{AW-NPA},
\begin{equation}
{\tilde S}_{\lambda\mu}^H(rel) = {2\over \gamma v}
{i^{\lambda+\mu} {\bar G}_{\lambda\mu}\over
\sqrt{(\lambda+\mu)!(\lambda-\mu)!}} \
\Bigl({\omega\over v}\Bigr)^\lambda \
K_\mu(\xi_b={\omega b\over \gamma v}).
\label{strel} 
\end{equation}
where ${\bar G}_{\lambda\mu}$ can be extracted from Ref. \cite{AW-NPA}.
The notation used here is such that ${\bar G}_{\lambda\mu}=1$ in
the non-relativistic limit, whereas Ref. \cite{AW-NPA} uses a different
normalization. For electric dipole and quadrupole excitations one finds that
\begin{equation}
{\bar G}_{10}={\bar G}_{20} = {\bar G}_{2\pm 2} = {1\over \gamma}, \ \ 
{\bar G}_{1\pm 1} = 1, \ \ {\bar G}_{2\pm 1} = 
{1\over 2}(1+{1\over\gamma^2}).
\end{equation}
It is seen that the relativistic effects on the Coulomb excitation 
enter into Eq. (\ref{strel}) through the factor 
$\gamma^{-1}{\widetilde G}_{\lambda\mu}$, and in the adiabatic
distance $\gamma v/\omega$ of the adiabaticity
parameter $\xi_b=\omega b/(\gamma v)$, which is the argument of the 
modified Bessel functions.

To complete the discussion of relativistic effects one should also 
specify the kinematics of Coulomb scattering at relativistic energies. 
This is done in appendix B. 
One of the reasons large relativistic effects have been observed 
is actually due to the difference between relativistic and 
non-relativistic kinematics, the main one being the determination
of the velocity from the beam energy, Eq. (B1). 
A relativistic effect in Coulomb scattering is the $\gamma$-factor which 
appears in the definition (B3) of half the distance of closest approach. 
This effect is commonly agreed upon \cite{AW-NPA,bert68}.
There is also a relativistic correction to the reduced mass,
Eq. (B2), which is less significant at intermediate energies,
and it is usually ignored \cite{AW-NPA,bert68}. However, for completeness,
it is better to keep it in the case of really high energies.
Finally, there are also relativistic effects in the transformation 
(B4) from the center-of-mass to laboratory scattering angles.

\subsection{Interpolating model} 

To accurately calculate the Coulomb excitation at intermediate beam 
energies it is important to include relativistic effects
and the effect of the Coulomb distortion on the relative trajectory
of projectile and target.
It may be difficult to derive in a general expression for the
Coulomb excitation amplitude because of the acceleration in a
Coulomb orbit. However, one can devise a formula which gives 
the correct expression in the non-relativistic regime for
a Coulomb orbit, and which reproduces the relativistic
expressions for a straight line trajectory in the
high-energy regime.  We shall see that the following expression,
\begin{equation}
S_{\lambda\mu}^H(rel) = {2\over \gamma v}
{i^{\lambda+\mu} \ {\bar G}_{\lambda\mu}\over
\sqrt{(\lambda+\mu)!(\lambda-\mu)!}} \
\Bigl({\omega\over v}\Bigr)^\lambda \
K_{\lambda\mu}^{eff}(b/a,\xi_a={\omega a\over \gamma v}),
\label{effective}
\end{equation}
serves the purpose of interpolating between the two energy regimes.

It is first noted that Eq. (\ref{effective}) is identical to
Eq. (\ref{nonrel}) if we insert $\gamma=1$. The low-energy
regime is therefore correctly described.
In the high-energy regime, we can assume that $b/a>>1$ and $\xi_a<<1$, 
which implies that the straight-line trajectory limit will be reached, 
$$
K_{\lambda\mu}^{eff}(b/a,\xi_a) 
\rightarrow K_\mu(\xi_b), \ \ \ {\rm where} \ \
\xi_b = \frac{b}{a} \xi_a, 
$$
according to the discussions in section IV.B and IV.C.
Since the value of $\xi_a$ in Eq. (\ref{effective}) is chosen 
as $\xi_a$ = $\omega a/(\gamma v)$ we obtain 
$$
\xi_b = \frac{b}{a} \times \frac{\omega a}{\gamma v} =
\frac{\omega b}{\gamma v},
\label{prescr}
$$
which is the correct adiabaticity parameter for a straight-line 
trajectory in the relativistic regime, according to Eq. (\ref{strel}), 
so the high-energy regime will also be described correctly.

\subsection{Total cross sections}

A great advantage of the straight-line trajectory approximation is 
that one can obtain analytic expressions for the total cross section, 
evaluated for all impact parameters larger that a certain minimum 
impact parameter $b_{0}$.
Thus one obtains \cite{AW-NPA}
\begin{equation}
\sigma_{\lambda} = 4\pi
\Bigl(\frac{Z_1e^2}{(2\lambda+1)\hbar v}\Bigr)^2
\ B(E\lambda) \sum_\mu 
\frac{4 {\bar G}_{\lambda\mu}^2} {(\lambda+\mu)!(\lambda-\mu)!} \
\Bigl({\omega\over v}\Bigr)^{2(\lambda-1)} \
g_\mu(\frac{\omega b_0}{\gamma v}),
\end{equation}
where
\begin{equation}
g_\mu(\xi) = 
2\pi \int_{\xi}^\infty \xi d\xi \ K_{\mu}^2(\xi)
=\pi\xi^2\Bigl[K_{\mu+1}(\xi)K_{\mu-1}(\xi) - K_\mu^2(\xi)\Bigr],
\end{equation}
according to Eq. (3.4) of Ref. \cite{AW-NPA}.
For dipole and quadrupole
excitations this yields
\begin{equation}
\sigma_{\lambda=1} =
16\pi \Bigl(\frac{Z_1e^2}{3\hbar v}\Bigr)^2 \ B(E1) \ 
\Bigl[\frac{1}{\gamma^2} \ g_0(\xi) +g_1(\xi)\Bigr],
\label{sdip}
\end{equation}
\begin{equation}
\sigma_{\lambda=2} =  \frac{4\pi}{3}
\Bigl(\frac{Z_1e^2}{5\hbar v}\Bigr)^2 \ B(E2) \
(\frac{\omega}{\gamma v})^2 \ 
\Bigl[3g_0(\xi) + g_1(\xi)\gamma^2(1+\gamma^{-2})^2 + g_2(\xi)\Bigr].
\label{squad}
\end{equation}

At intermediate and high energies, where one can assume that $\xi_b<< 1$, 
one obtains the following simple asymptotic expression for the quadrupole 
excitation cross section,
\begin{equation}
\sigma_{\lambda=2} =  \frac{1}{3} \
\Bigl(\frac{4\pi Z_1e^2}{5\hbar v b_0}\Bigr)^2 \ B(E2).
\label{simquad}
\end{equation}
This expression gives a surprisingly good estimate of the cross 
section at high energies and it provides a simple way of testing
more elaborate numerical calculations.  
The expression (\ref{simquad}) shows that the high-energy cross 
section is insensitive to the excitation energy.
One can also conclude that relativistic effects are not dramatic for 
E2 transitions because all of the $\gamma$ factors that appear in Eq. 
(\ref{squad}) disappear when one takes the limit $\xi_b\rightarrow$0. 

We shall see in the next section that the large relativistic effects,
which have been pointed out in the literature, are primarily caused by 
plotting the cross sections as a function of the beam energy $T$ per nucleon.
The cross section (\ref{simquad}), which is proportional to $v^{-2}$,
will therefore be very sensitive to whether one uses non-relativistic 
($v=\sqrt{2T/m}$) or relativistic kinematics (Eq. (B1)) to determine
the velocity.

For completeness it is noted that some relativistic effects do survive 
in the dipole cross section, Eq. (\ref{sdip}), when one takes the limit 
$\xi_b\rightarrow$0 in the dipole cross section, Eq. (\ref{sdip}),
\begin{equation}
\sigma_{\lambda=1} =
\Bigl(\frac{4\pi Z_1e^2}{3\hbar v}\Bigr)^2 \ B(E1) \ 
\Bigl[2\ln(\frac{1.123 \gamma v}{\omega b_0}) - (\frac{v}{c})^2\Bigr].
\label{simdip}
\end{equation}
This expression shows the well-know logarithmic dependence on $\gamma$. 


\section{Applications}

Two examples of calculated cross sections are shown in Fig. 3,
namely, for the dipole excitation of $^{11}$Be to the low-lying $1/2^-$ 
state, and the quadrupole excitation of $^{42}$S to the lowest $2^+$ 
state, respectively.
The cross sections are shown as functions of the beam energy.
In both cases the minimum impact parameter was set to $b_0$ = 14 fm
and the target was Au. The excitation energy and multipole strength
of the two transitions are given in Table I, which will be discussed
below.

The upper thick dashed curves in Fig. 3 show the relativistic 
Coulomb excitation cross section for straight-line trajectories.
The solid curves are based on Coulomb trajectories and make use of
the expression, Eq. (\ref{effective}), for the interpolating model.
The lower, thin dashed curves in Fig. 3 are the cross sections one 
obtains by inserting $\xi_a=\omega a/v$ into the expression, 
Eq. (\ref{effective}), for the interpolating model. 
This is seen to be a poor approximation.  Inserting instead 
$\xi_a=\omega a/(\gamma v)$ (the thick solid curves) one obtains 
a smooth transition from the low-energy to the high-energy theory.

\subsection{Comparison to experiments}

A number of Coulomb excitation experiments have been performed at
intermediate energies and some of them are quoted in Table I.
Those considered here are of interest  because sufficient 
experimental information was provided to repeat the analysis.
The measured cross sections, $\sigma_{exp}$, can be compared
to the straight-line trajectory approximation, $\sigma_{Strl}$, 
and to the cross section, $\sigma_{Coul}$, obtained from 
the interpolating model, Eq.  (\ref{effective}).
It is seen that the latter model performs very well in comparison
to the measured cross sections. 
The deviations are insignificant in comparison to the experimental 
uncertainty.

It should be emphasized that most of the B(E2) values quoted in Table I 
were extracted from the data using the relativistic straight-line 
trajectory approximation with some corrections included for the
Coulomb distortion of the trajectory. The good agreement 
between $\sigma_{Coul}$ and the measured cross sections 
therefore shows that the previous analyses were very reasonable.
The results of the straight-line trajectory approximation,
$\sigma_{Strl}$, are quoted in the last column. They are
typically 5 to 10\% higher than the measured values.

The ratio of measured and calculated cross sections 
is illustrated in Fig. 4.
The solid circles show the ratio $\sigma_{exp}/\sigma_{Coul}$
with respect to the interpolating relativistic Coulomb excitation
cross section. The average ratio is about 2\% less than 1.
The triangles in Fig. 4 show the experimental ratio 
$\sigma_{exp}/\sigma_{Strl}$ to the relativistic straight-line
trajectory calculation. Here the deviation from 1 is much larger. 
The deviation from the solid points reflects the significance of 
the Coulomb distortion of the trajectory. 
It amounts to about 3 - 9\%.
The largest deviations from the solid points occur in the
low energy experiments, runs no. 4-8 (see Table I.) 


There are two examples in Table I where the B(E2) values are 
known from other sources, namely, $^{24}$Mg and $^{26}$Mg. 
The calculated cross sections, $\sigma_{Coul}$, are also in 
these cases consistent with the measured cross sections. 
This is very fortunate because the uncertainties are small 
(about 5\% on the measured cross sections and 3\% on the B(E2) values).
These two measurements therefore provide an independent test 
of the interpolating model, Eq. (\ref{effective}).

The last example on a comparison with data is the Coulomb 
excitation of $^{46}$Ar \cite{ar46}, which was measured for a 
range of maximum acceptance angles. 
The measurements are compared to two calculations in Fig. 5,
namely, the straight-line and the `interpolated' Coulomb 
trajectory calculation, Eq. (\ref{effective}).
Both calculations are in good agreement with the data
because the experimental uncertainties are much larger 
than the difference between the two calculations.

\subsection{Comparison to other methods}

The relativistic description for straight-line trajectories developed 
by Winther and Alder \cite{AW-NPA} was based on the Li\'enard-Wiechert 
potential.  An different method was used by Aleixo and Bertulani
\cite{aleixo} who calculated the retardation effects explicitly for 
Coulomb trajectories. 
The latter approach was used by Bertulani et al. \cite{bert68,bertpl} 
to investigate the effects of relativity and Coulomb distortions at 
intermediate energies in much the same way it is done here. 
It is therefore of interest to compare Eq. (\ref{effective}) to the 
predictions of the more elaborate theory.

An example of a comparison of calculated cross sections is shown
in Fig. 6 for the Coulomb excitation of $^{46}$Ar to the $2^+$ on 
a Pb target.  The input is the same as used in producing table 2 of 
Ref. \cite{bertpl}, and it  was assumed \cite{bertcom} that the 
distance of closest approach for a Coulomb trajectory,
$r_0(b) = a+\sqrt{a^2+b^2}$ according to Eqs. (\ref{coultr}) and 
(\ref{eps}), has the minimum value 
$r_{min}$ = $R_1+R_2=1.2(A_1^{1/3}+A_2^{1/3})$.
The solid line is the cross section obtained from the interpolating 
model, Eq. (\ref{effective}). The upper dashed curve is the result 
for straight line trajectories with minimum impact parameter 
$b_{min}=\sqrt{r_{min}(r_{min}-2a})$. 
The lower dashed curve is the 
non-relativistic cross section for Coulomb trajectories.
It is seen that the solid curve interpolates smoothly between the
non-relativistic Coulomb trajectory calculation at low energy and the 
relativistic straight-line trajectory calculation at high energy.

The solid points in Fig. 6A are based on the cross sections published
in table 2 of Ref. \cite{bertpl}. They have here been multiplied by 
the factor 1.44 because a factor of $e^2$ was unfortunately omitted
\cite{bertcom} in the calculations of Ref. \cite{bertpl}.
The results are presented in Fig. 6B in terms of the ratio 
to the interpolating relativistic Coulomb trajectory calculation,
Eq. (\ref{effective}). The average value of the solid points is
close to one (actually 1.005 $\pm$ 0.002, to be precise)
which shows that the two theories are in excellent agreement.
This may not be so surprising because both theories approach the 
relativistic straight-line trajectory approximation at high energy 
and the non-relativistic Coulomb trajectory calculation at low energy.

The lower dashed curve in Fig. 6B shows that relativistic effects 
are enormous at 500 MeV/nucleon.  It is interesting that the large 
relativistic effects
have a very simple explanation.
Thus, according to the asymptotic expression, Eq. (\ref{simquad}), 
the quadrupole excitation cross section is proportional to $v^{-2}$
at high energies.
In the non-relativistic calculation this implies
$\sigma_2^{\rm NR}\propto m/(2T)$. In the relativistic calculation
one obtains instead $\sigma_2^{\rm rel}\propto$
$(m+T)^2/[T(2m+T)]$, according to Eq. (B1) of appendix B.
The ratio of the two cross sections is therefore
\begin{equation}
\frac{\sigma_2^{\rm NR}}{\sigma_2^{\rm rel}} \approx
\frac{m(2m+T)}{2(m+T)^2}.
\label{vrel} 
\end{equation}
(There is also a difference in the minimum impact parameter
in the relativistic and non-relativistic calculations but the 
effect is small.) 
The expression (\ref{vrel}) is illustrated in Fig. 6B by the triangles 
which explain quite accurately the behavior of the non-relativistic 
calculation at high energy. 

\subsection{Significance of relativistic effects}

Another way to illustrate the effects of relativity is to recalculate the 
cross sections shown in Fig. 3 assuming that $\gamma=1$ everywhere in the 
underlying equations, Eqs.  (\ref{strel}-\ref{effective}). 
The velocity $v$, which appears explicitly in these equations, will be 
determined from the relativistic expression, Eq. (B1), in order to avoid 
the trivial relativistic effect described by Eq. (\ref{vrel}).
The results of such calculations show that the cross sections for the 
excitation of the $2^+$ state in $^{42}$S, which were shown in Fig. 3,
change by less than 0.5\%, both for the straight-line trajectory 
calculation (\ref{strel}) and also for the interpolated Coulomb trajectory 
model, Eq. (\ref{effective}). 

Similar calculations performed for the dipole excitation of $^{11}$Be 
(which were shown in Fig. 3) change the cross section by less than 1\% 
at energies below 200 MeV/nucleon. The change is about 10\% at 1 GeV/u but 
that is not so surprising because the asymptotic dipole cross section, 
Eq.  (\ref{simdip}), does contain a $\gamma$ factor, whereas the
asymptotic quadrupole cross section, Eq. (\ref{simquad}), does not.

It is concluded that the relativistic effects in the Coulomb excitation 
of nuclei are large at intermediate and high energies but most 
of the effect is trivial and can easily be avoided by using the correct
relativistic expression to determine the velocity from the beam
energy.  In the analysis of measurements it is also important to use 
relativistic kinematics when converting scattering angles into impact 
parameters. The non-trivial relativistic effects on the Coulomb 
quadrupole excitation, on the other hand, are surprisingly small. 
 
\subsection{Significance of Coulomb distortion}

Let us finally take a look at how the straight-line trajectory 
approximation can be corrected for Coulomb distortions.
Two examples are shown in Fig. 7, namely, the Coulomb 
excitation to the $2^+$ state of $^{16}$N with excitation energy 
$E_x$ = 0.12 MeV (A), and also to the $2^+$ state of $^{54}$Ni with 
excitation energy $E_x$ = 1.4 MeV (B). The B(E2) values were taken 
from table 2 of Ref. \cite{bertpl}.
The results are shown in terms of the ratio to the interpolating, 
relativistic Coulomb excitation cross section, Eq. (\ref{effective}).
The calculations were again performed with the minimum distance of 
closest approach $r_{min}= 1.2 (A_1^{1/3}+A_2^{1/3})$.

The comparison with Ref. \cite{bertpl} is illustrated in Fig. 7 by 
the solid points. The cross sections from table 2 of Ref. \cite{bertpl}
were again multiplied with the factor 1.44. It is seen that the
solid points are very close to 1 for the excitation of $^{54}$Ni.
There are some discrepancies for $^{16}$N, where the average ratio 
is 1.028 $\pm$ 0.012.  It is therefore concluded that the two theories
of relativistic Coulomb excitation agree within a few percent.

The thick solid curves in Fig. 7 show the ratio for the relativistic 
straight-line trajectory approximation, and the dashed curves show the 
results of various ways to correct this approximation for Coulomb distortions. 
The top dashed curves show the straight-line approximation multiplied 
by the factor $\exp(-\pi\xi_a)$.
This factor has a large effect for the heavier nucleus $^{54}$Ni with
the relatively large excitation energy but it has essentially no effect 
for the lighter nucleus $^{16}$N with the small excitation energy.

The lowest dashed curves in Fig. 7 show the straight-line trajectory 
result one obtains by replacing the minimum impact parameter
$b_{min}$ by the effective value $b_{eff}$ = $b_{min}+\pi a/2$. 
This approximation was justified in Ref. \cite{AW-NPA} for
large impact parameters, where the excitation probability 
falls off exponentially as $\exp(-2\omega b/(\gamma v))$.
Thus by multiplying the excitation probability for a straight
line trajectory, $P_{Strl}$, with the factor $\exp(-\pi\xi_a)$
one obtains approximately,
$$
P_{Strl} \ \exp(-\pi\xi_a) \propto
\exp(-\frac{2\omega b}{\gamma v}) \ 
\exp(-\pi\frac{\omega a}{\gamma v}) 
= \exp[-\frac{2\omega}{\gamma v}(b+\frac{\pi a}{2})].
$$
This argument does not always apply to the Coulomb excitation 
of low-lying states at intermediate energies because the minimum 
impact parameter is usually much smaller than the adiabatic distance 
$\gamma v/\omega$. Using the effective minimum impact parameter has 
a very large effect on the calculated cross section. 
It produces a ratio in Fig. 7 that is almost as far below 1 as the 
ratio for the pure straight line trajectory calculation is above 1.  
The approximation is therefore not very useful for quadrupole excitations.
It works apparently better for dipole excitations, as discussed in 
connection with Figs. 2 and 3, but that will not be discussed here.
 
The second lowest curves in Fig. 7A and 7B are the results one obtains 
by choosing the effective minimum impact parameter,
$b_{eff}=r_{min}$, which is the minimum distance of closest 
approach that is used in the relativistic Coulomb trajectory calculation. 
This approximation was recommended by Goldberg \cite{goldberg},
and it is evidently the best choice out of the four examples of 
approximations shown in Fig. 7. 

In the analysis of an actual experiment the cross section ratio 
discussed above would usually be closer to one because one would always 
choose a small acceptance angle (i.~e., a large minimum impact parameter) 
in order to avoid the influence of nuclear and higher-order processes.

\section{Conclusions}

A model has been devised for including relativistic effects in calculations
of the Coulomb excitation cross section at intermediate energies. 
The model interpolates smoothly between the theory of non-relativistic 
Coulomb excitation at low energy and the relativistic, straight-line 
trajectory approximation high energy. 
The results that were obtained with this model compare very well with the 
calculations performed by Bertulani et al., who included the relativistic 
retardation effects explicitly for Coulomb trajectories.

It was demonstrated that the large relativistic effects, which have
been pointed out in the literature, are mainly caused by 
comparing calculations that are based on 
a relativistic and a non-relativistic velocity, respectively.
The non-trivial relativistic effects, which are beyond the simple 
relativistic kinematics of two-body scattering, are surprisingly 
small for quadrupole excitations.

The Coulomb distortion, which is responsible for the deviation between
the straight line trajectory approximation and calculations that
are based on a Coulomb trajectory, can also have a very large effect.
However, the effects of the Coulomb distortion are usually suppressed 
by the experimental conditions and simple corrections can be made to 
improve the accuracy of the straight line trajectory approximation.

The proposed interpolating model reproduces fairly well the analyses
that have been performed previously, primarily of quadrupole excitation 
experiments, at beam energies in the range of 30 to 70 MeV/nucleon.
The average deviation is only a few percent.  
The model also reproduces the measured cross sections in the few cases 
where the quadrupole excitation strength is known accurately from other 
sources. 

The good agreement with the previous analyses is partly due to the 
experimental conditions, which suppress the effects of the Coulomb 
distortion, and partly to the fact that some corrections for the 
distortion were made in the analyses.
However, if high precision Coulomb excitation experiments were pursued, 
it would be necessary to treat the Coulomb distortion more accurately
in the analysis. It is believed that the interpolating Coulomb excitation 
model proposed here would provide a sufficiently accurate description.

{\bf Acknowledgments.}
This work was supported by the U.S. Department of Energy,
Office of Nuclear Physics, under Contract No. DE-AC02-06CH11357.

\section{Appendix A: Straight line trajectory}

In coordinate system A, a straight line trajectory has the 
coordinates $x(t)=b$, where $b$ is the (constant) impact parameter, 
and $y(t)=vt$. Using the dimensionless integration variable $s=vt/b$, 
the orbital integral, Eq. (\ref{slma}), is therefore
$$
{\tilde S}_{\lambda\mu}^A = d_{\mu0}^\lambda({\pi\over 2}) \
{1\over vb^\lambda} \int_{-\infty}^\infty ds \ e^{i\xi_bs} \
{(1+is)^\mu\over (1+s^2)^{(\lambda+\mu+1)/2}}
$$
$$
= d_{\mu0}^\lambda({\pi\over 2}) \
{1\over vb^\lambda} \Bigl(1+{d\over d\xi_b}\Bigr)^\mu 
\int_{-\infty}^\infty ds \ {\cos(\xi_bs)
\over (1+s^2)^{(\lambda+\mu+1)/2}},
\eqno(A1)
$$
where $\xi_b=\omega b/v$ is the adiabaticity parameter.
The tilde on ${\tilde S}_{\lambda\mu}^A$ is a reminder that we are
using the straight-line approximation.
An analytic expression for the integral is given in the book 
by Gragshteyn and Ryzhik \cite{gradshteyn}, Eqs. 8.432 no. 5, 
and one obtains
$$
{\tilde S}_{\lambda\mu}^A = d_{\mu0}^\lambda({\pi\over 2})
{2\over vb^\lambda} {1\over (2n-1)!!}
\Bigl(1+{d\over d\xi_b}\Bigr)^\mu \xi_b^n K_n(\xi_b),
\label{strlas}
\eqno(A2)
$$
which is expressed in terms of modified Bessel functions of order
$n=(\lambda+\mu)/2$. Here $\lambda+\mu$ is even as mentioned earlier
so $n$ is an integer. We can express the result in a form similar
to Eq. (\ref{cltr}) with
$$
{\tilde I}_{\lambda,\pm\mu} =
\frac{2}{(2n-1)!!} \ \bigl(\frac{a}{b}\bigr)^\lambda \
(1\pm \frac{d}{d\xi_b})^\mu \ \xi_b^n K_n(\xi_b).
\label{strlai}
\eqno(A3)
$$
To evaluate this expression one can make use of the relations:
$\frac{d}{dx} (x^n K_n(x)) = - x^n K_{n-1}(x)$.
For dipole and quadrupole excitations one obtains
$$
{\tilde I}_{1,\pm 1} = 2\xi_a \ 
\Bigl [K_1(\xi_b) \mp K_0(\xi_b) \Bigr], 
\label{strla1}
\eqno(A4)
$$
$$
{\tilde I}_{20} = \xi_a^2 \Bigl[K_2(\xi_b) - K_0(\xi_b)\Bigr]
= 2\ (\frac{a}{b})^2 \ \xi_b \ K_1(\xi_b),
\label{I20strl}
\eqno(A5)
$$
$$
{\tilde I}_{2\pm 2} = \frac{1}{3} \xi_a^2 \Bigl[K_2(\xi_b) \mp 
4K_1(\xi_b)+3K_0(\xi_b)\Bigr], 
\label{strla2}
\eqno(A6)
$$
where the relation $K_2(x)=K_0(x)+2/xK_1(x)$ has been used in Eq. (A5).

To improve the straight-line approximation, one can multiply the 
results, Eqs. (A4-A6), by the factor 
$\exp(-\pi\xi_a/2)$, according to Winther and Alder \cite{AW-NPA}.
In fact, the exact analytic expression, which has been obtained 
for a Coulomb trajectory in the case of dipole excitations, supports 
this suggestion; see Eq. (12) in appendix H of Ref. \cite{AW-book}.
We will later on investigate how good this improvement and other
approximations are for quadrupole excitations.

\section{Appendix B: Relativistic kinematics}

Here we specify the expressions that are used to calculate the
relativistic Coulomb scattering. Most of them are taken from
Jackson's book \cite{jackson}.
First of all, the kinetic energy T of 
the projectile is commonly given in units of MeV/nucleon so the $\gamma$ 
factor and the beam velocity $v$ can be obtained from
$$
\gamma m = m +T, \ \ \ \beta = v/c= 
\frac{\sqrt{T(2m+T)}}{m+T},\eqno(B1) 
$$
where $m$ = 931.5 MeV is the nucleon mass (using the notation $c$=1.)

The masses of projectile and target are denoted by $M_1$ and $M_2$,
and the total energy in the center of mass system is (Jackson (12.31)) 
$$
E' = \sqrt{M_1^2+M_2^2 + 2\gamma M_1M_2}.
$$
where $\gamma M_1$ is the laboratory energy of the projectile.
Moreover, the energy and momentum of the projectile in the
center of mass system are (Jackson, Eqs. (12.31-34))
$$
E_1' = \frac{M_1^2+E_1M_2}{E'} = \frac{M_1(M_1+\gamma M_2)}{E'},
$$
$$
p_1' = \frac{M_2}{E'} p = \frac{M_1M_2}{E'} \ \gamma v.
$$
where $p=\gamma M_1 v$ is the momentum of the projectile in
the laboratory frame.
Note that the non-relativistic reduced mass $M_1M_2/(M_1+M_2)$ has 
been replaced in the last expression by the relativistic reduced mass,
$$
M_0 = \frac{M_1M_2}{E'} = 
\frac{M_1M_2}{\sqrt{M_1^2+M_2^2 + 2\gamma M_1M_2}}.\eqno(B2)
$$

Rutherford's scattering formula in the center-of-mass system is 
derived from the transverse momentum transfer in elastic Coulomb
scattering, estimated in the straight line approximation by 
$$
\Delta p_{\perp} = p_1' \ \sin(\theta) =
\frac{2 Z_1Z_2e^2}{vb},
$$
This is a reasonable approximation in high-energy forward-angle
scattering but to match the non-relativistic expression,
$\tan(\theta/2)=a/b$, one should consider the Coulomb distortion 
of the trajectory.  This would give a factor of $\cos^2(\theta/2)$ 
on the right-hand side, so we obtain
$$
\tan(\theta/2) = \frac{Z_1Z_2 e^2}{p_1' v b} =
\frac{Z_1Z_2 e^2}{\gamma M_0 v^2 b}.
$$
Thus we recover the usual scattering formula,
$\tan(\theta/2)=a/b$, but the definition of $a$, Eq. (\ref{headon}), 
must be replaced by
$$
a = \frac{Z_1Z_2 e^2}{\gamma M_0 \beta^2}.\eqno(B3)
$$
There are two corrections compared to Eq. (4). One is the factor 
$1/\gamma$, which is commonly considered. The other is the relativistic 
reduced mass $M_0$, which is often replaced by the non-relativistic 
value.

The scattering angle in the laboratory frame is determined by
(see Jackson (12.50))
$$
\tan(\phi) = \frac{\sin(\theta)}{\gamma_{c.m.}(\cos(\theta) + \alpha)},
\eqno(B4)
$$
where $\gamma_{c.m.}=(\gamma M_1+M_2)/E'$ (Jackson (12.35)), and 
$$
\alpha= \frac{M_1}{M_2} \ \frac{M_1+\gamma M_2}{\gamma M_1+M_2},
\eqno(B5)$$
according to Jackson (12.54) for elastic scattering.  It is seen that 
the transformation from the c.m. to the laboratory system reduces to 
the usual expression for $\gamma\rightarrow 1$.

\begin{table}
\caption{Cross sections for the 
quadrupole excitation ($\lambda=2$) of different nuclei on a Au or Bi target, 
and the dipole excitation ($\lambda=1$) of $^{11}$Be (last row).
The experimental conditions are from the quoted references.
The $T$ is the beam energy per nucleon at mid-target and
$\phi_{max}^{Lab}$ ($\theta_{max}^{c.m}$) is the maximum laboratory 
(center-of-mass) acceptance angle.
The adopted B(E2) values \cite{NDS} for $^{24}$Mg and $^{26}$Mg are also shown.
The the last two columns show the calculated cross sections for
the relativistic Coulomb ($\sigma_{Coul}$)
and straight-line ($\sigma_{Strl}$) trajectories.}
\begin{tabular} {|c|c|c|c|c|c|c|c|c|}
\colrule
 run &          & E$_x$ & B(E$\lambda$) & T & $\phi_{max}^{Lab}$ & $\sigma_{exp}$ & 
$\sigma_{Coul}$ &  $\sigma_{Strl}$ \\
 no. &  Nucleus & (MeV) & (e$^2$fm$^{2\lambda}$) & (MeV/nucleon) & (deg) & (mb) & (mb) & (mb) \\ 
\colrule
 0 & $^{24}$Mg+Au \cite{simg} & 
   1.3687& 467(28) & 36   & $\theta_{max}^{c.m.}\leq$ 4.48 & 78.7(48) &  81.7 &  88.0 \\
 & adopted \cite{NDS} &       & 436(10) &    &               &      &  76.3(18) &  82.2 \\
 0 & $^{26}$Si+Au \cite{simg} & 
                1.7959& 336(33) & 41.8 & $\theta_{max}^{c.m.}\leq$ 4.48   & 55.8(55) &  56.3 &  61.0 \\
\colrule
1 & $^{26}$Mg+Bi \cite{mg} & 1.8087 & 315 & 67   & 2.38 &  44(2)  &  45.9 &  47.8 \\
  & adopted \cite{NDS}     &       & 307(9) &      &      &      &  44.7(13)&  46.6 \\
2 & $^{32}$Mg+Au \cite{mg} & 0.885 & 447 & 71.2 & 2.26 &  91(10) &  92.7 &  95.9 \\
3 & $^{34}$Mg+Bi \cite{mg} & 0.659 & 541 & 67   & 2.38 & 126(22) & 130.1 &  134.9 \\
\colrule
4 & $^{38}$S+Au \cite{sar}   & 1.292 & 235 & 34.6 & 4.10 &  59(7)  &  60.1 &  64.7 \\
5 & $^{40}$S+Au \cite{sar}   & 0.891 & 334 & 35.3 & 4.10 &  94(9)  &  96.9 & 103.4 \\
6 & $^{42}$S+Au \cite{sar}   & 0.890 & 397 & 36.6 & 4.10 & 128(19) & 131.1 & 139.9 \\
\colrule
7 & $^{44}$Ar+Au \cite{sar}  & 1.144 & 345 & 30.9 & 4.10 &  81(9)  &  83.0 &  89.6 \\
8 & $^{46}$Ar+Au \cite{sar}  & 1.554 & 196 & 32.8 & 4.10 &  53(10) &  53.6 &  58.3 \\
9 & $^{46}$Ar+Au \cite{ar46} & 1.554 & 212 & 73.2 & 2.90 &  68(8 ) &  68.6 &  71.9 \\
\colrule 
 & $^{11}$Be+Au \cite{be11} & 0.32  &0.079& 57.6 & 3.80 & 244(31) & 246.1 & 247.2 \\
\colrule
\end{tabular}
\end{table}

\begin{figure}
\includegraphics [width = 12cm]{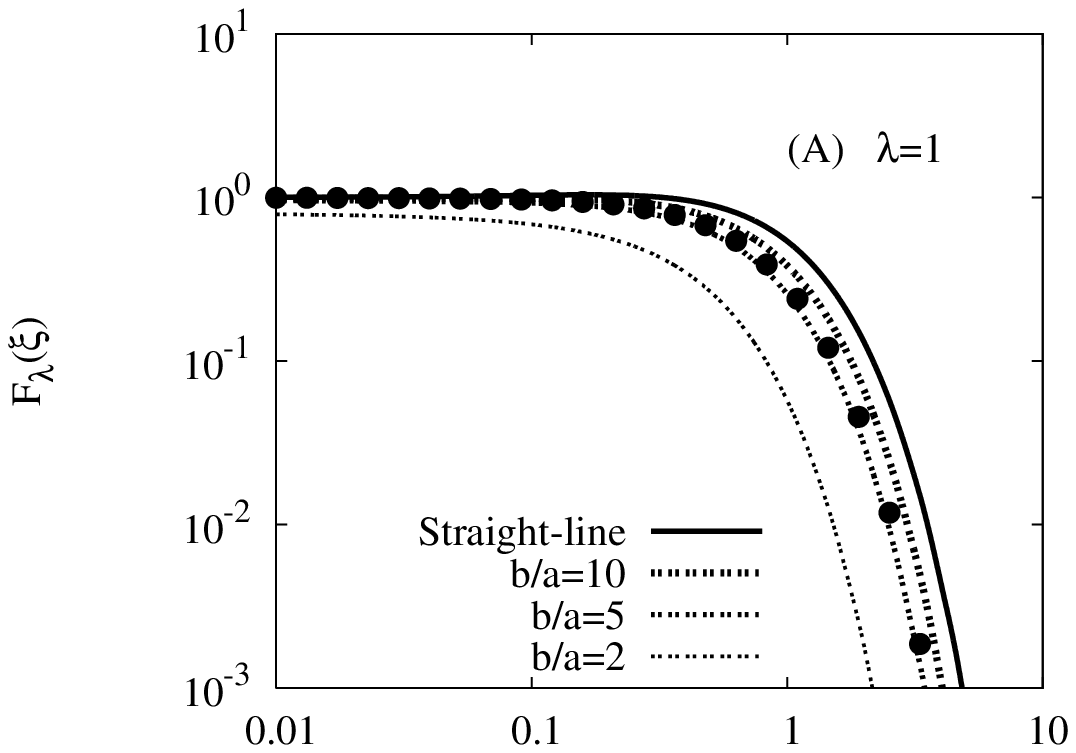}
\includegraphics [width = 12cm]{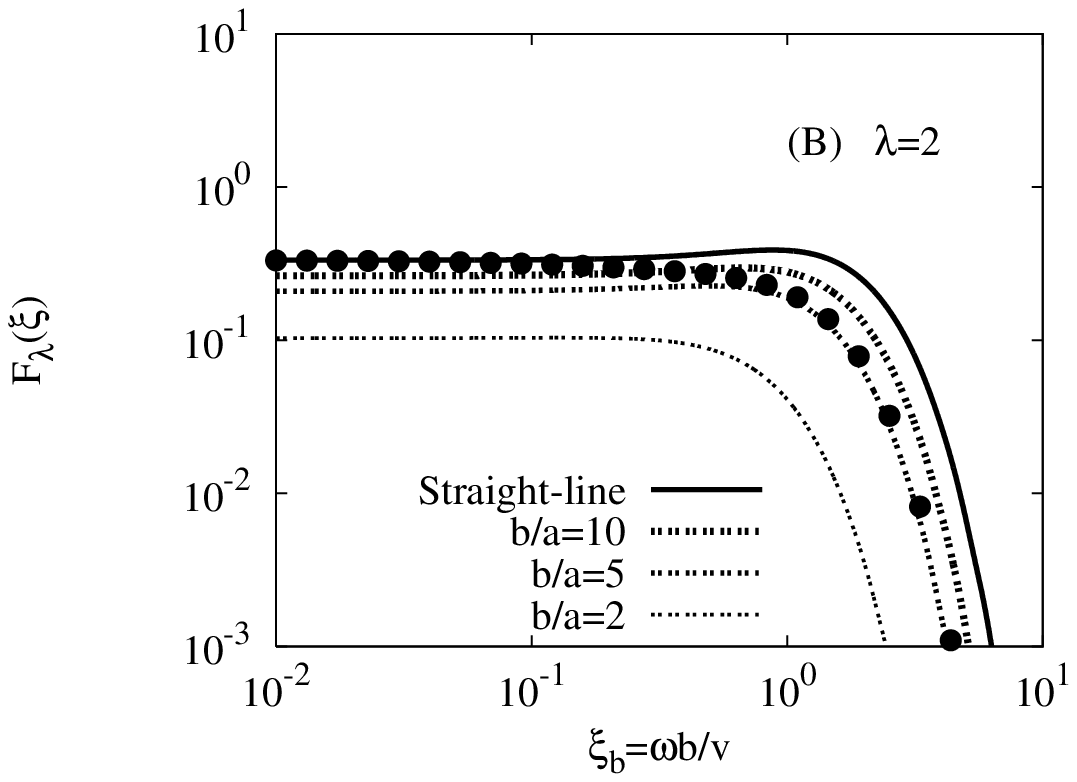}
\caption{The adiabaticity functions $F_\lambda(b/a,\xi)$ are
illustrated for $\lambda$ = 1 and 2.
The top solid curves are the results for a straight line trajectory,
Eq. (\ref{fstrl}). The dashed curves are obtained from Coulomb trajectories,
Eq. (\ref{fcoul}), using the indicated values of $b/a$. 
The solid points were obtained by multiplying the straight line trajectory 
with $exp(-\pi\xi_a)$ = $exp(-\pi\frac{a}{b}\xi_b)$ for $a/b$=5.}
\label{total}
\end{figure} 

\begin{figure}
\includegraphics [width = 11cm]{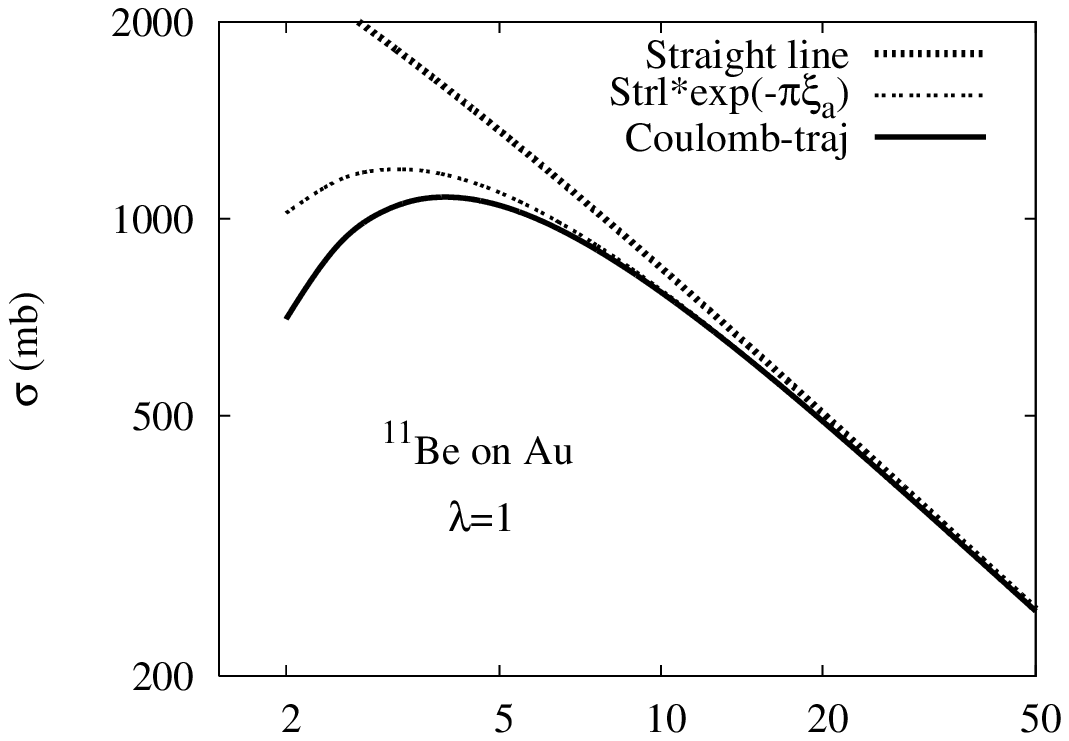}
\includegraphics [width = 11cm]{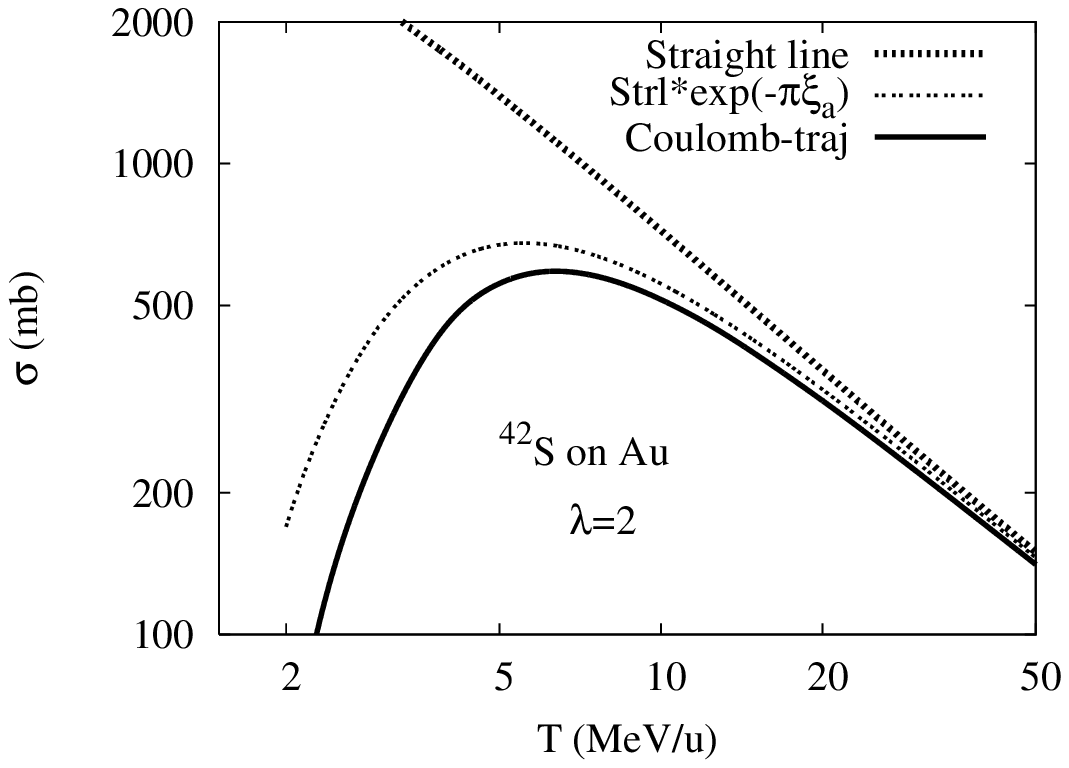}
\caption{Energy dependence of the non-relativistic cross sections
for the dipole excitation of $^{11}$Be and the quadrupole excitation
of $^{42}$S discussed in the text.  The target is gold, and the 
minimum impact parameter was set to 14 fm in both cases.
The upper thick dashed curves show the straight-line trajectory 
approximation, whereas the thin dashed curves have been corrected 
with the factor $\exp(-\pi\xi_a)$. 
The solid curves are the results for Coulomb trajectories.}
\label{be11au}
\end{figure} 

\begin{figure}
\includegraphics [width = 11cm]{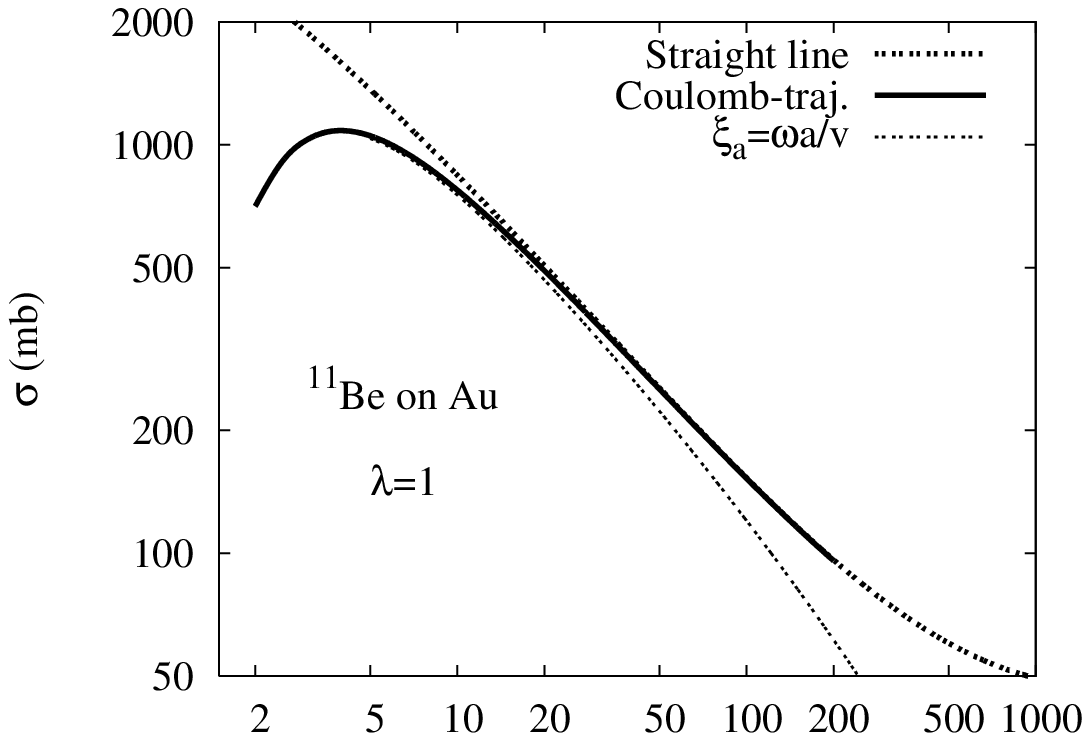}
\includegraphics [width = 11cm]{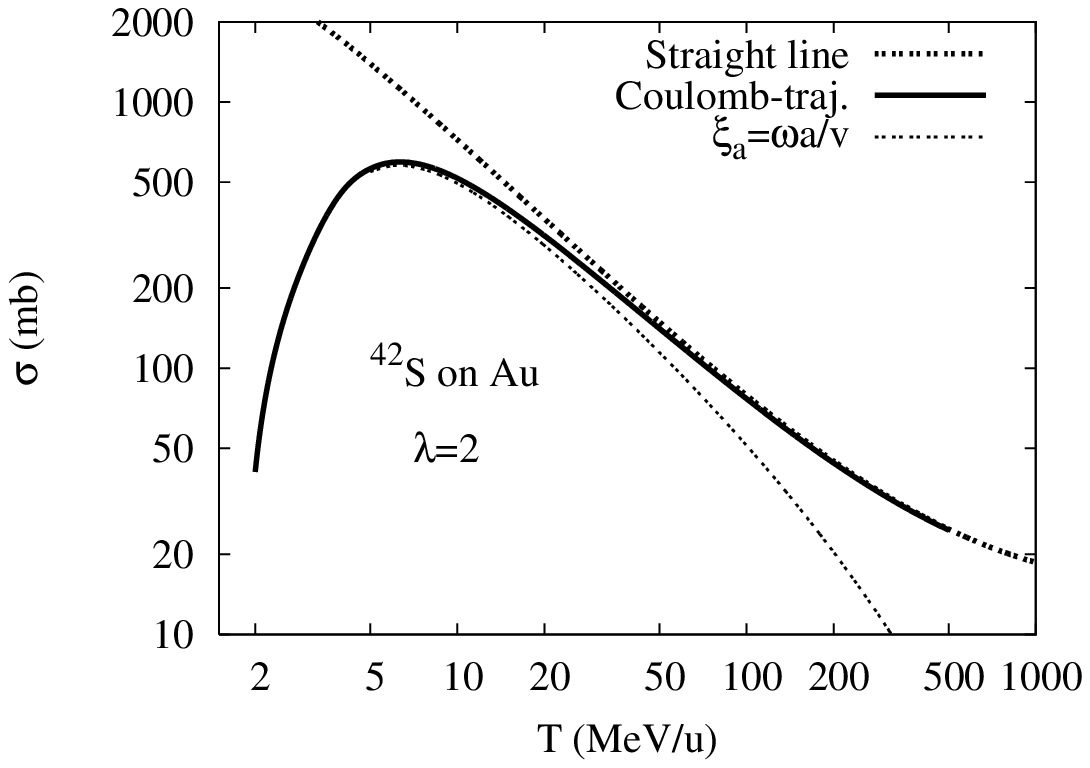}
\caption{Energy dependence of the relativistic cross sections
for the dipole excitation of $^{11}$Be and the quadrupole 
excitation of $^{42}$S discussed in the text.
The target is Au, and the minimum impact parameter was set
to 14 fm in both cases.
The thick dashed curves show the straight-line trajectory 
approximation, Eqs. (\ref{sdip},\ref{squad}). 
The solid curves show the interpolating, relativistic Coulomb 
excitation cross section, Eq. (\ref{effective}). 
The thin dashed curves are also based on Eq. 
(\ref{effective}) but use $\xi_a=\omega a/v$
in the second argument of $K_{\lambda\mu}^{eff}(b/a,\xi_a)$.}
\label{be11aunr}
\end{figure} 

\begin{figure}
\includegraphics [width = 11cm]{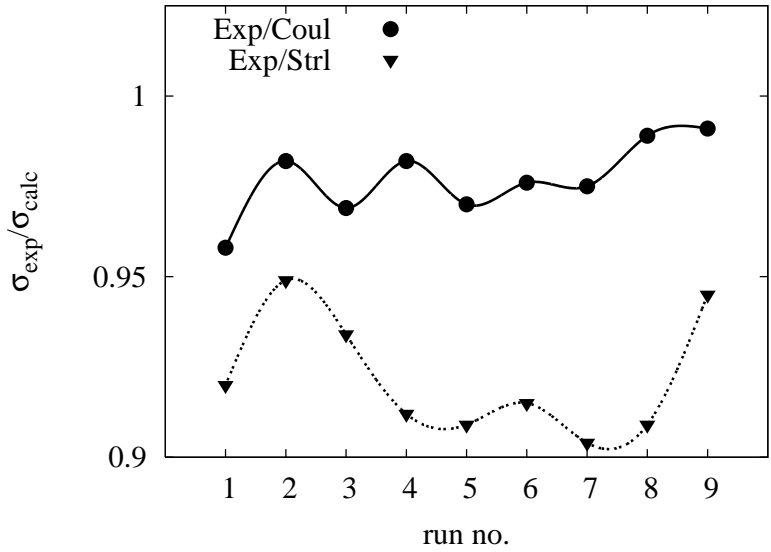}
\caption{Ratio of measured and calculated Coulomb excitation cross 
sections for the different runs shown in Table I. Shown are the results 
for the interpolated relativistic Coulomb excitation (solid points) and 
the straight-line trajectory calculations
(triangles).}
\label{runs}
\end{figure} 

\begin{figure}
\includegraphics [width = 12cm]{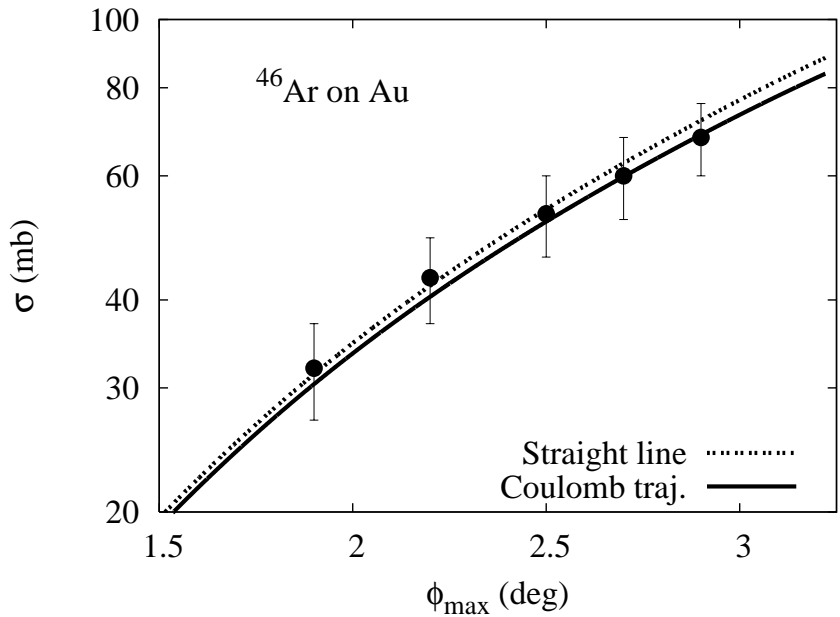}
\caption{Cross section for the Coulomb excitation of the $2^+$ state 
in $^{46}$Ar on a Au target at 73.2 MeV/nucleon as a function of the 
maximum laboratory acceptance angle $\phi_{max}$. The relativistic
straight-line and Coulomb trajectory calculations are compared to 
the data \cite{ar46}.}
\label{ar46b}
\end{figure} 

\begin{figure}
\includegraphics [width = 11cm]{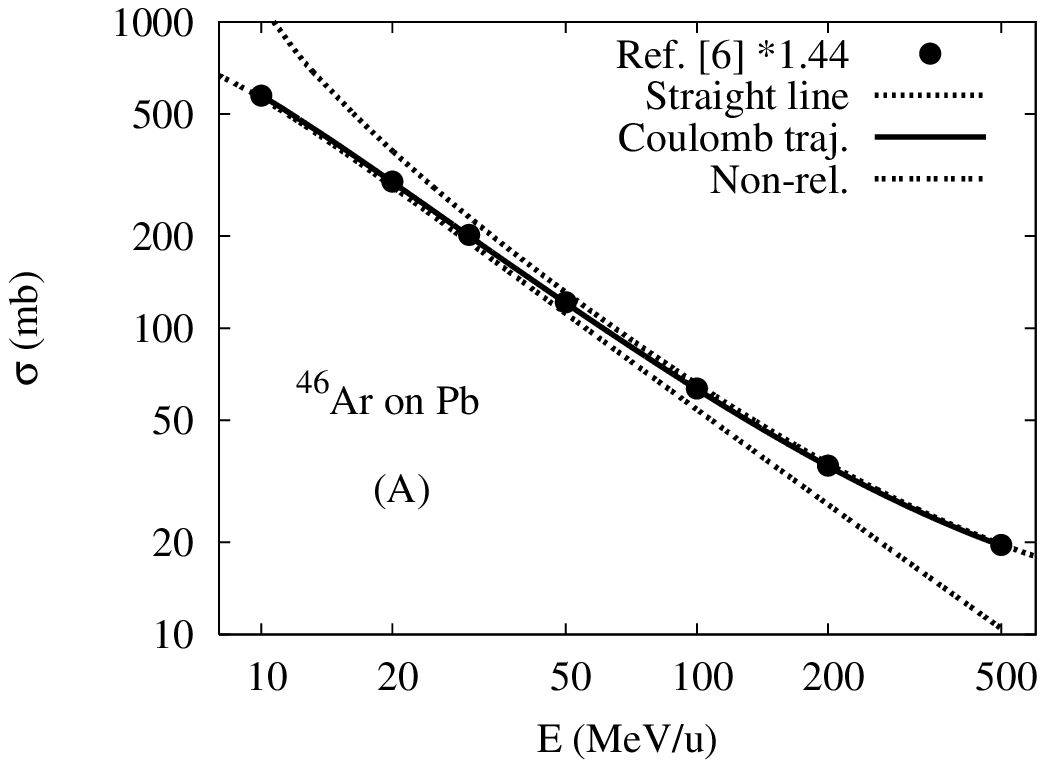}
\includegraphics [width = 11cm]{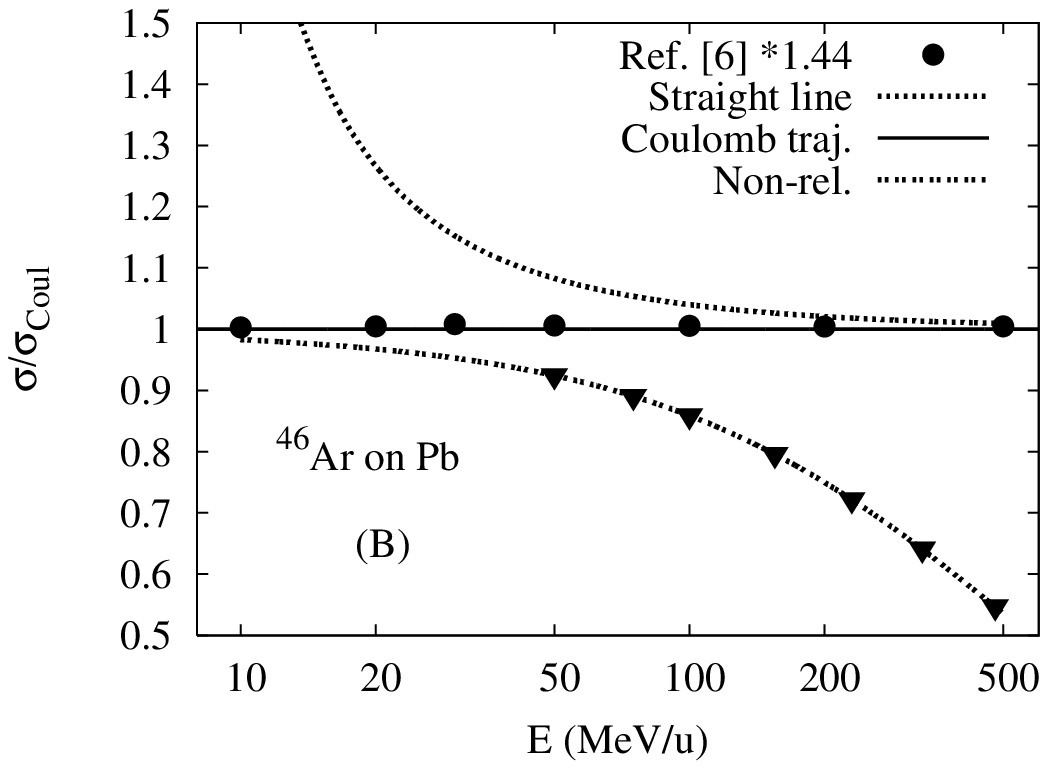}
\caption{(A):
cross sections for the $2^+$ excitation of $^{46}$Ar on a Pb target.
The minimum distance of closest approach was set to 11.41 fm.
The relativistic straight-line (upper dashed curve) and 
Coulomb trajectory calculations (solid curve) are compared to the 
non-relativistic Coulomb trajectory calculation (lower dashed curve), 
and to the results of Ref. \cite{bertpl} (solid points)
which have been multiplied by 1.44. 
(B): ratios of the cross sections in (A)
to the relativistic Coulomb trajectory calculation.
The triangles show the estimate, Eq. (\ref{vrel}).}
\label{ar46pb}
\end{figure} 

\begin{figure}
\includegraphics [width = 11cm]{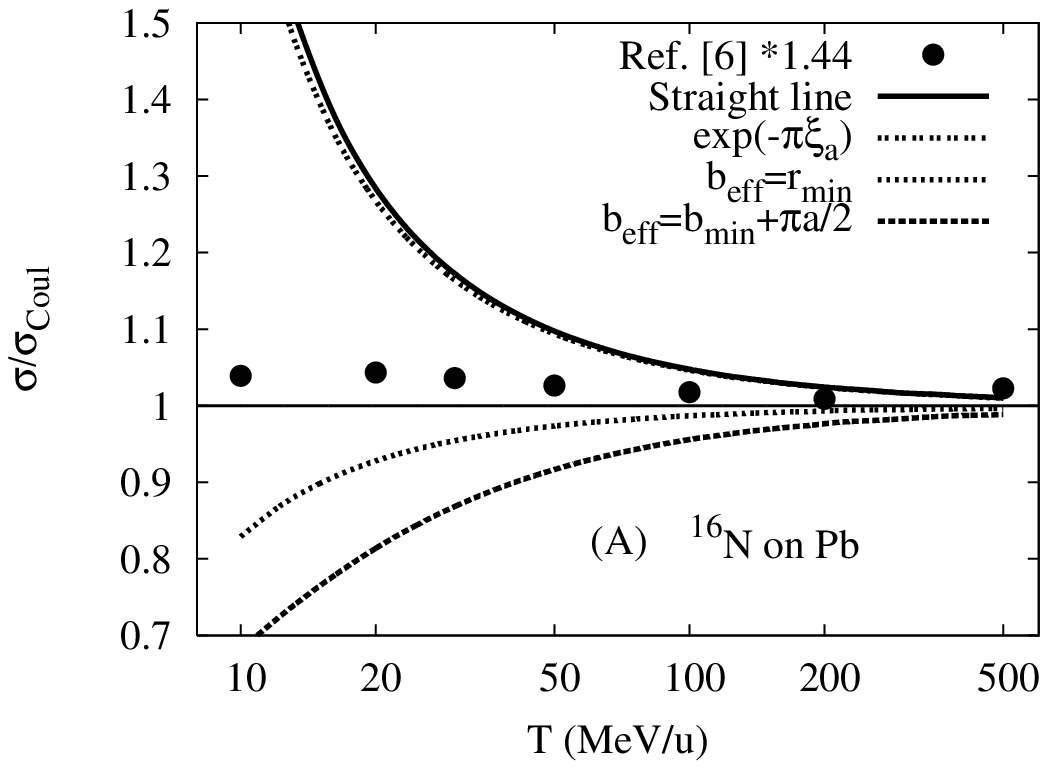}
\includegraphics [width = 11cm]{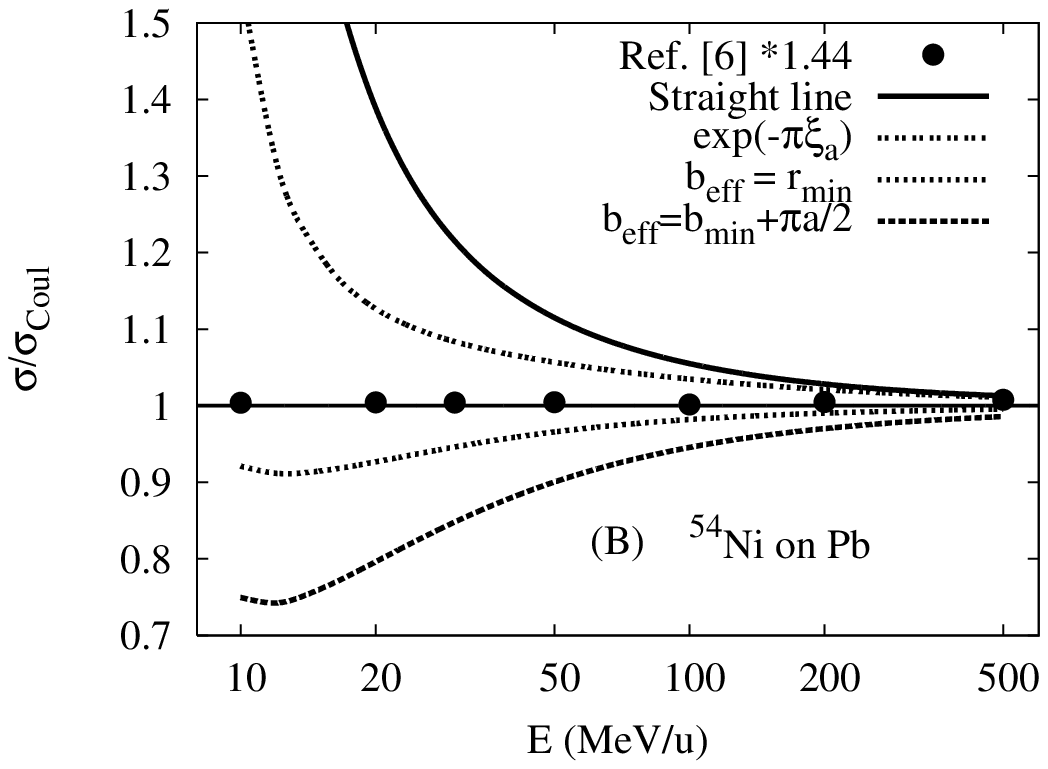}
\caption{
Ratios of different cross sections to the relativistic Coulomb trajectory 
calculation. Results are shown for the excitation of the $2^+$ state in
$^{16}$N and $^{54}$Ni, respectively, on a Pb target.
The thick solid curves are the ratios for the straight line trajectory
approximation. The dashed curves show various ways of correcting the 
straight-line approximation, such as multiplying with the factor 
$\exp(-\pi\xi_a)$, or using the effective minimum impact parameters 
discussed in the text. The solid points are the ratios for the cross 
sections given in table 2 of Ref. \cite{bertpl} multiplied by 1.44.}
\label{ratios}
\end{figure} 


\begin{thebibliography}{99}
\bibitem{AW-book} K. Alder and A. Winther, {\it Electromagnetic Excitation}
(North-Holland, NY, 1975).
\bibitem{AW-NPA} A. Winther and K. Alder, Nucl. Phys. {\bf A319}, 518 (1979).
\bibitem{glass} T. Glasmacher, Annual Rev. Nucl. Part. Sci. {\bf 48}, 1 (1998).
\bibitem{aleixo} A. N. F. Aleixo and C. A. Bertulani, Nucl. Phys. {\bf A505},
448 (1989).
\bibitem{bert68} C. A. Bertulani, A. E. Stuchbery, T. J. Mertzimekis, and 
A. D. Davies, Phys. Rev. C {\bf 68}, 044609 (2003).
\bibitem{bertpl} C. A. Bertulani, G. Cardella, M. D. Napoli, G. Raciti,
and E. Rapisarda, Phys. Lett. B {\bf 650}, 233 (2007).
\bibitem{scheit} H. Scheit, A. Gade, Th. Glasmacher, T. Motobayashi,
Phys. Lett. B {\bf 659}, 515 (2007). 
\bibitem{AW56} K. Alder, A. Bohr, T. Huus, B. Mottelson, and A. Winther,
               Rev. Mod. Phys. {\bf 28}, 432 (1956). 
\bibitem{simg} P. D. Cottle {\it et al}., Phys. Rev. C {\bf 64}, 057304 (2001). 
\bibitem{NDS} Evaluated Nuclear Structure Data Files (ENSDF), National 
     Nuclear Data Center, Brookhaven National Laboratory,
     http//:www.nndc.bnl.gov/
\bibitem{mg} J. A. Church {\it et al}., Phys. Rev. C {\bf 72}, 054320 (2005). 
\bibitem{sar} H. Scheit {\it et al}., Phys. Rev. Lett {\bf 77}, 3967 (1996). 
\bibitem{ar46} A. Gade {\it et al}., Phys. Rev. C {\bf 68}, 014302 (2003). 
\bibitem{be11} M. Fauerbach {\it et al}., Phys. Rev. C {\bf 56}, R1-4 (1997).
\bibitem{bertcom} C. A. Bertulani (private communications).
\bibitem{goldberg} A. Goldberg, Nucl. Phys. {\bf A240}, 636 (1984).
\bibitem{gradshteyn} I. S. Gradshteyn and I. M. Ryzhik,
{\it Table of Integrals, Series and Products} (Academic Press, NY, 1965).
\bibitem{jackson} J. D. Jackson, {\it Classical Electrodynamics} (Wiley, NY, 1967).
\end{thebibliography}
\end{document}